\documentclass[graybox]{svmult}

\usepackage{mathptmx}       
\usepackage{helvet}         
\usepackage{courier}        
\usepackage{type1cm}        
\usepackage{makeidx}         
\usepackage{graphicx}        
\usepackage{multicol}        
\usepackage[bottom]{footmisc}
\usepackage{url}

\makeindex             

\begin{document}

\title*{A Bayesian Approach to Gravitational Lens Model Selection}
\author{Ir\`ene Balm\`es}
\institute{Ir\`ene Balm\`es \at Laboratoire Univers et Th\'eories (LUTh), UMR~8102 CNRS, Observatoire de Paris, Universit\'e Paris Diderot, 5 place Jules Janssen, 92190 Meudon, France\\ irene.balmes@obspm.fr}

\maketitle

\abstract{Strong gravitational lenses are unique cosmological probes. These produce multiple images of a single source. Whether a single galaxy, a group or a cluster, extracting cosmologically relevant information requires an accurate modeling of the lens mass distribution. A variety of models are available to this purpose, nevertheless discrimination between them as primarely relied on the quality of fit without accounting for the size of the prior model parameter space. This is a problem of model selection that we address in the Bayesian statistics framework by evaluating Bayes' factors. Using simple test cases, we show that the assumption of more complicate lens models may not be justified given the level of accuracy of the available data.}
  
    Images produced by strong gravitational lenses result of different light-paths. If the source behind the lens has a variable luminosity, this will manifest with a time delay between the images.
This time delay $\varDelta t$ depends on the gravitational potential of the lens, and the underlying cosmological model. Therefore, we can derive constraints on cosmological parameters (in particular $H_0$), provided a lens model is assumed. Hence, lens modeling as well as accurate measurements capable of discriminating between models are critical to the study of time delays.

      We aim to tackle this problem from the point of view of Bayesian model selection analysis (see e.g.~\cite{Trotta}). A large number of lens models have been proposed in a vast literature. Given the fact that observables are limited to the position of the images, their time delay and flux ratio, we restrict our analysis to simple examples characterized by a few parameters. In particular we consider two models for lenses with two images, so called ``double'' lenses (for a review on lensing, see~\cite{Kochanek}).        
      \begin{enumerate}
        \item{Power-law model:} assume a density profile $\varrho \propto r^{-n}$, with $n$ a free parameter. For $n=2$, it describes an isothermal lens. In order to assess the dependence on the prior parameter interval we assume two different priors: $0 < n < 3$ (large) and $1 < n < 3$ (small).

        \item{Power-law model with external shear:} assume the previous model with the addition of shear accounting for environmental effect on the lens. This adds two parameters: the strength of the shear $\gamma$, and its direction. Expected values for the shear vary up to $\gamma \simeq 0.1$, therefore we assume three different priors on $\gamma$: $\gamma < 0.1$, $< 0.2$ and $< 0.5$ respectively. This allow us to test the shear strength up to nearly unrealistic values.  
      \end{enumerate}

We performed a likelihood data analysis for a sample of lenses and inferred the Bayes' factor for model 1 and 2 under different priors.

\begin{figure}[t]
  \includegraphics[width=\textwidth]{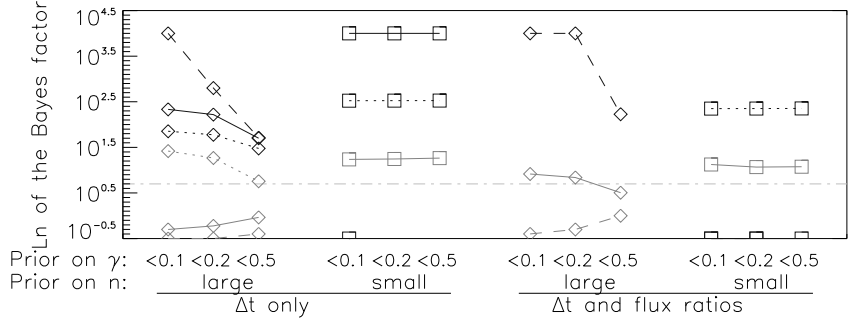}
  \caption{Bayes'  factor between model~1 and~2, with different priors. Above the dot-dashed line, the evidence in favor of model~1 is strong. Each color represents a different lens}
\label{evidence}  
\end{figure}

Results are summarized in Fig.~\ref{evidence}. Large Bayes' factors favor the simpler model, model~1. Above a certain threshold (dot-dashed line), the evidence in favor of model~1 is considered strong. In the following, we highlight a few relevant aspects.

  \begin{itemize}
    \item{Effect of the prior on $n$:}
    The lens data set is mainly composed of galaxies, which we expect to be nearly isothermal. Nevertheless, our analysis shows that a large fraction of our sample is accurately described by model~1 if $0<n<1$.
    \item{Effect of the prior on $\gamma$:}
    In more than half of the cases, allowing higher (unrealistic) shear strength does not change the Bayes' factor. This illustrates the effect of the Occam's razor term in the Bayes' factor: a wider range for a parameter is bound to give a better fit, but this is balanced against a penalty factor. 
    \item{Effect of the flux ratios:}
    Time delays depend on the gravitational potential of the lens, whereas flux ratios depend on its second derivative. Furthermore, they are subject to a number of local phenomena that do not affect time delays. As a result, flux ratios require more complex models than time delays. This is consistent with our findings in Fig.~\ref{evidence}: indeed, adding flux ratios as a constraint leads to having less lenses accurately described by model~2.
  \end{itemize}

\begin{acknowledgement}
P.-S.~Corasaniti provided helpful advice on both this work and this paper. 
I.~Balm\`es is supported by a scolarship of the "Minist\'ere de l'\'Education Nationale, de la Recherche et de la Technologie" (MENRT).
\end{acknowledgement}

\bibliographystyle{plain}
\bibliography{Balmes_procSCMAV}

\begin{thebibliography}{1}

\bibitem{Kochanek}
C.~S. {Kochanek}.
\newblock {Part 2: Strong gravitational lensing}.
\newblock In {G.~Meylan, P.~Jetzer, P.~North, P.~Schneider, C.~S.~Kochanek, \&
  J.~Wambsganss}, editor, {\em Saas-Fee Advanced Course 33: Gravitational
  Lensing: Strong, Weak and Micro}, pages 91--268, 2006.

\bibitem{Trotta}
R.~{Trotta}.
\newblock {Bayes in the sky: Bayesian inference and model selection in
  cosmology}.
\newblock {\em Contemporary Physics}, 49:71--104, March 2008.

\end{thebibliography}

\end{document}